# COOL DWARFS IN WIDE MULTIPLE SYSTEMS

## PAPER 5: NEW ASTROMETRY OF 54 WIDE PAIRS WITH M DWARFS

*By Miriam Cortés-Contreras,*
*Departamento de Astrofísica, Universidad Complutense de Madrid*

*José A. Caballero*
*Centro de Astrobiología (CSIC-INTA), Madrid, Spain*

*and David Montes*
*Departamento de Astrofísica, Universidad Complutense de Madrid*

We investigate the membership in double, triple or higher-order-multiplicity systems of 54 pairs with at least one bright M dwarf in the solar neighbourhood. These M dwarfs are potential targets of radial-velocity surveys for exoplanets. We measure angular separations and position angles from optical images taken with TCP and CAMELOT at the IAC80 telescope at the Observatorio del Teide, and complement them with our measurements on photographic plate digitizations. We also use data in the Washington Double Star Catalogue and other bibliographic sources. We confirm the physical binding of 52 multiple systems, for which we comprehensively compile, derive and provide basic astrophysical parameters in a homogeneous way (spectral types, heliocentric distances, projected physical separations, individual masses, estimated orbital periods, binding energies). Of the 52 systems, 38 are double, 11 are triple and three are quadruple with a variety of architectures. Four systems contain white dwarfs, six systems display variations of position angle larger than 12 deg (1/30 orbit) on a scale of decades and seven systems are located at less than 10 pc. We provide new information, or correct published data, of the most remarkable multiple systems and identify some of them for high-resolution imaging and spectroscopic follow-up.

### *Introduction*

There are thousands of cool main-sequence stars in known pairs, many of which have projected physical separations of hundreds or thousands of astronomical units (Paper 1 of this series and references therein[1]). The existence of cool dwarfs in wide multiple systems helps in the investigation of their formation and evolution, especially if the other component is a Sun-like star (useful for, *e.g.,* metallicity studies), a white dwarf (useful for, *e.g.,* nuclear-age determination) or even an identical M dwarf with different rotation period or X-ray emission (useful for, e.g., comparative magnetic-braking models). Besides, the nearest systems have correspondingly very wide angular separations and, therefore, can be easily be resolved from the ground with standard imagers and telescopes of moderate size. In some cases, and with long enough astrometric monitoring, one can study the relative movement of the two stars in the pair without the aid of high-resolution-imaging devices (speckle, adaptive optics, lucky imaging), which is of great help for determining dynamical masses.

In this work, we investigate in detail several dozen wide pairs with M dwarfs with a threefold objective: confirming their true common proper-motion (and, thus, membership in a physical system), homogenously characterising a large set of pairs, many of which have never been investigated astrometrically in detail, and searching for remarkable multiple wide systems (*i.e.,* pair candidates for which an astrometric orbit can be calculated, systems with secondaries without proper spectral-type determinations, the most fragile pairs at the boundary of disruption by the Galactic gravitational field, or triple –and quadruple– systems).

*Observations*

Originally, the observation programme that led to the results presented here was aimed at imaging stars selected as potential targets for upcoming near-infrared radial-velocity exoplanet surveys (such as HPF[2], SPIRou[3] or, especially, CARMENES[4]). Such stars must be the least-active, brightest, latest-type M dwarfs[5] with no companions at less than 5 arcsec (a separation at which the flux of any visual or physical companion could affect the radial-velocity measurement of the main target[6]). We used the 0.82-m IAC80 telescope at the Observatorio del Teide for imaging 103 fields with at least one such M dwarf. The targets were selected from a large list of potential targets because of poor UCAC3[7] optical photometry (at the time of preparing the observations, UCAC4[8] had not yet been published), presence of nearby sources in virtual-observatory images that may prevent accurate spectroscopic follow-up, or even membership in hypothetical common proper-motion pairs or multiple systems of unknown status. The observations were performed in service mode from mid-2012 to the beginning of 2013.

Owing to a problem in one of the instruments at the IAC80, we observed 99 fields with the Tromsø CCD Photometer (TCP[9]) and four fields only with the Cámara Mejorada Ligera del Observatorio del Teide (CAMELOT[10]). TCP and CAMELOT provide fields of view and pixel scales of 9.2 × 9.2 arcmin$^2$ and 0.537 arcsec/pixel, and 10.37 × 10.37 arcmin$^2$ and 0.304 arcsec/pixel, respectively (Fig. 1). In both cases, we used the Johnson *R* filter in short (~60 s) and long (~300 s) exposures, which increased our dynamic range in the innermost arcseconds close to the targets. We also tuned the exposure times to *try* to avoid saturation.

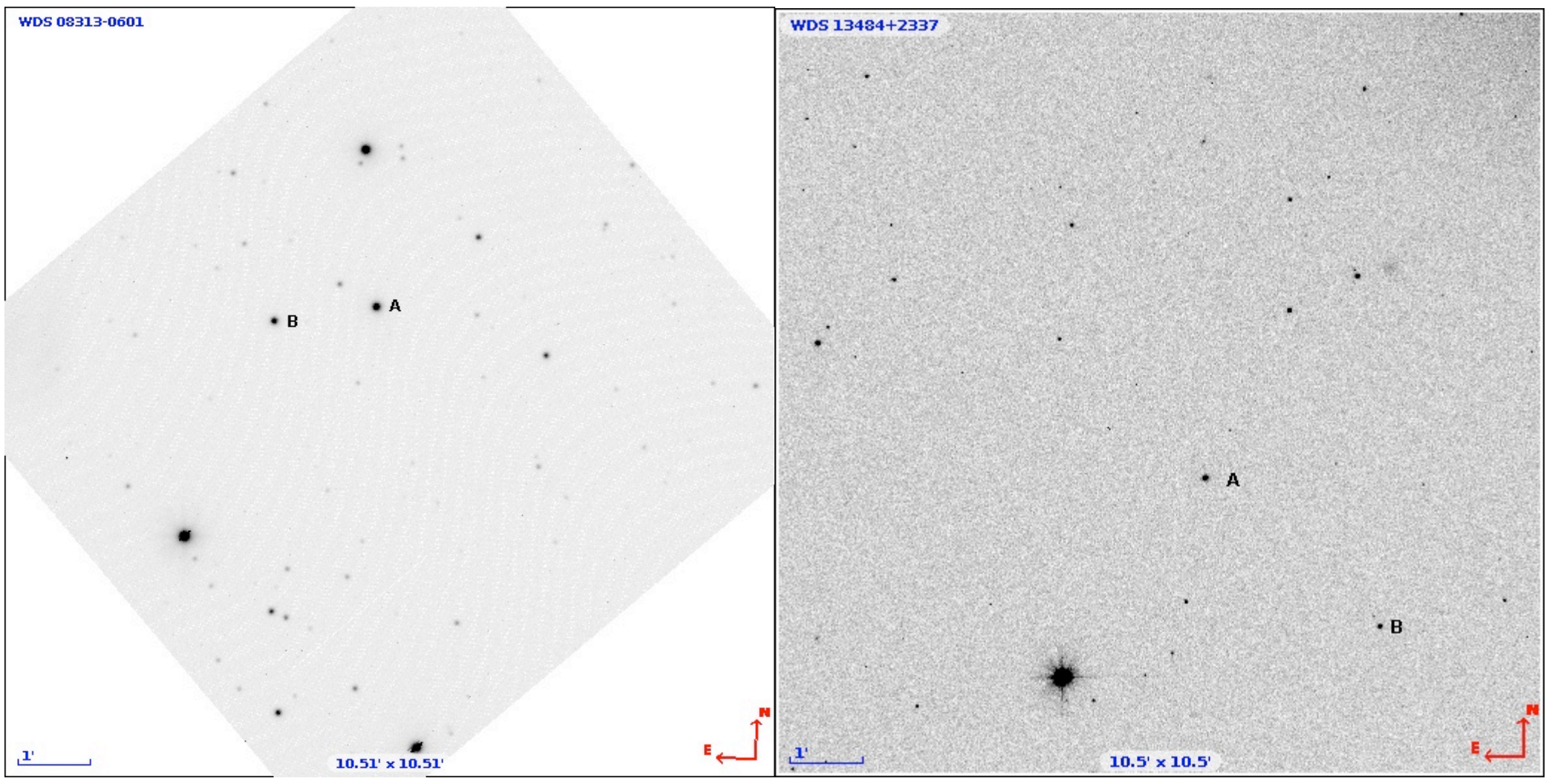

FIG. 1

Representative images taken with TCP (*left*, WDS 08313−0601) and CAMELOT (*right*, WDS 13484+2337). A and B components are tagged in both images. Exposure times were 60 s each. The 50.85-degree tilt of the TCP field of view is obvious.

Of the 103 fields, 56 had at least one known candidate companion to the M dwarf, either brighter or fainter than our nominal target. The reader must not deduce that the binary frequency of M dwarfs is higher than 50% because our sample was biased towards stars in binaries and triple systems (multiplicity, either in the form of very close binaries of equal brightness or systems with very bright primaries, tends to affect photometric measurements).

Unfortunately, the images for three systems were not useful because of intense saturation of the primaries. In particular, we also observed but discarded from the analysis: WDS 03575−0110 (BU 543: BD−01 565 AB), WDS 04153−0739 (STF 518: o Eri AB-C), and WDS 20408+1956 (LDS 1045: GJ 797 AB). As a result, our final sample consisted of 53 double or triple systems with at least one M dwarf. Of them, 52 were observed with TCP and one (WDS 13484+2337) with CAMELOT.

*Astrometric analysis*

After applying bias and flat-field corrections, we measured angular separations $\rho$ and positions angles $\theta$ on the best processed images of each pair (in general, the long-duration images for systems with all their components faint, the short-duration images for the rest). Both $\rho$ and $\theta$ were measured taking into account the pixel size and the detector orientation of TCP and CAMELOT (Fig. 1). We used the *imexam* task within the IRAF environment or the *distance* task within the Aladin sky atlas[11] for measuring on-CCD separations between stellar photocentroids depending on the quality of the match of their point-spread functions to a Gaussian profile (the brightest stars had guyot-like PSFs).

Uncertainties were calculated by error propagation. The Aladin measure errors were higher than IRAF's due to the uncertainty in the by-eye estimation of the stellar photocentroids (about 0.4 arcsec; *i.e.,* a bit less than one TCP pixel), but still acceptable for our purposes.

Table I summarizes our results. Since WDS 02457+4456 was supposed to be triple, we tabulate 54 pairs in 53 systems. We provide their Washington Double Star (WDS[12]) Catalogue identifications, discoverer codes, recommended names of both primary and secondary (note our restricted use of the letters A, B, C, a and b), and IAC80 angular separations, position angles, and observation epochs in Julian years. Measured angular separations range from 5.05 arcsec to 4.65 arcmin. The $\rho$ and $\theta$ values tabulated for WDS 07397+3328, marked in parenthesis, must *not* be used for astrometric purposes (see below).

We studied a further nine pairs for which we had controversial information on common proper motion, such as very different catalogued values of proper motion, or of $\rho$ or $\theta$ at the first and last epochs as tabulated by WDS or with respect to our own measurements. We applied the same virtual-observatory astrometric methodology as by Caballero[13] and the rest of papers in this series. In particular, we used our own astrometry on SuperCOSMOS digitizations of digital sky survey POSS-I and II (first and second Palomar Observatory Sky Survey) and UKST (United Kingdom Schmidt Telescope) photographic plates[14], the astrometry provided by the wide-field and all-sky catalogues 2MASS[15], CMC14[16], GSC 1.3 and 2.2[17], SDSS-DR9[18], and *WISE*[19], and our observations with IAC80. The epochs of observation, $\rho$ and $\theta$ values, and origin of every measurement (76 in total, most of which are new) are provided in Table II.

Two of our pairs, shown in italics in Table I, turned out to be optical systems (*i.e.,* not physical binaries). As a result, we do not compute average $\rho$ and $\theta$ values for them in Table II. One of the visual binaries is WDS 02457+4456 AB (GIC 34), which is formed by the red dwarfs G 78−4 (M0.5 V) and G 78−3 (M5.0 V). In spite of being 4.5 subtypes cooler, the hypothetical secondary is only about 0.5 mag fainter than the primary. Neither unresolved high-order multiplicity, metallicity, nor inflation[20], could explain such an overbrightness. Furthermore, in his master thesis, Dorda[21] had already reported a significant difference in proper motion between the two stars (from *Hipparcos*[22] and PPMXL[23]). Our astrometric follow-up showed a clear linear variation of both $\rho$ and $\theta$ between 1951 December and 2012 September, with an amplitude of 8.4 arcsec in angular separation. Indeed, their proper motions, measured by us, were found to be quite different (G 78−4: +411.8±4.5, −124.6±1.2 mas yr$^{-1}$; G 78−3: +276.4±3.6, −167.8±2.7 mas yr$^{-1}$). The simplest explanation for these observables is that G 78−3 is in the foreground at an estimated distance of 13.6±1.1 pc, while G 78−4 is located further at 23.1±1.2 pc. However, the latter forms the true common proper-motion pair WDS 02457+4456 AC (LDS 5393) together with the ultracool dwarf LP 197−48 (see below). To sum up, WDS 02457+4456 is not a triple system, but only a double one. However, and quite interestingly, G 78−4 and G 78−3 are, with angular separation only about 1.5 arcmin, the closest unrelated M dwarfs in the CARMENES input catalogue[6].

The other viusal binary is WDS 10585−1046 (LDS 4041), which is formed by BD−10 3166, a super-metal-rich K0 V star at 64–80 pc that hosts an exoplanet candidate[24,25,26], and LP 731−076, an M4 V star at about only 11 pc[25,27]. In this case, our astrometric study showed deviations of 3.6 arcsec in $\rho$ and 6.9 deg in $\theta$ in almost six decades, with different proper motions (especially in declination: BD−10 3166: −181.1±3.0, −5.2±1.8 mas $yr^{-1}$; LP 731−076: −187.7±3.9, −77.2±0.9 mas $yr^{-1}$).

The case of the system WDS 07397+3328 (LDS 3755) is so pure chance that it deserves a comic strip (Fig. 2). The primary is G 090−016 (M2.0 V), which at the POSS-I epoch (1955 February) was aligned with a background star, and the secondary is LP 256−44, which *today* is now aligned with the same background star. We had to take special care to disentangle the secondary from the interloper, of intermediate brightness between the primary and the secondary, in the photographic plates obtained in the 1990s. Actually, our IAC80 astrometric measurement is strongly affected by the background star ($\rho$ and $\theta$ values in Tables I and II –in parenthesis– correspond instead to the angular separation and position angle of the background star with respect to the primary). Probably because of this unfortunate alignment, it has never been possible to take a spectrum of the secondary.

WDS tabulates an angular separation of 38.0 arcsec for the first astrometric epoch of WDS 09288−0722 (GIC 87), in 1964. This value contrasted with the most recent measures by 2MASS and by ourselves, which lie at about 35.7 arcsec. From Table II, the constancy of the ten astrometric epochs between 1953 December and 2012 December led us to conclude that Ross 439 A and B do form a common proper-motion pair separated by 35.73±0.16 arcsec (the Giclas' measure in 1964[28] is only listed in the WDS as being to the nearest degree and nearest arcsecond, so it is just an estimated value).

Two more pairs have components with discordant tabulated proper motions for primaries and secondaries. In one, WDS 12123+5429 (VYS 5) is a pair of two bright early red dwarfs: BD+55 1519A (M0.0 V) and BD+55 1519B (M3.0 V). In the other, WDS 23294+4128 (GIC 193) is composed of G 190−28 (M3.5 V) and G 190−27 AB (M4.0 V+m5: V). As shown in Table III, some tabulated proper motions were incorrect (LSPM's for VYS 5, USNO-B1 and PPMXL's for GIC 193), while the two pairs actually move together through space.

The remaining dubious pairs for which we confirmed the constancy of their angular separations and position angles and, therefore, their common proper motions with an astrometric study were WDS 03398+3328, WDS 20446+0854, and WDS 22058+6539. They will be discussed next.

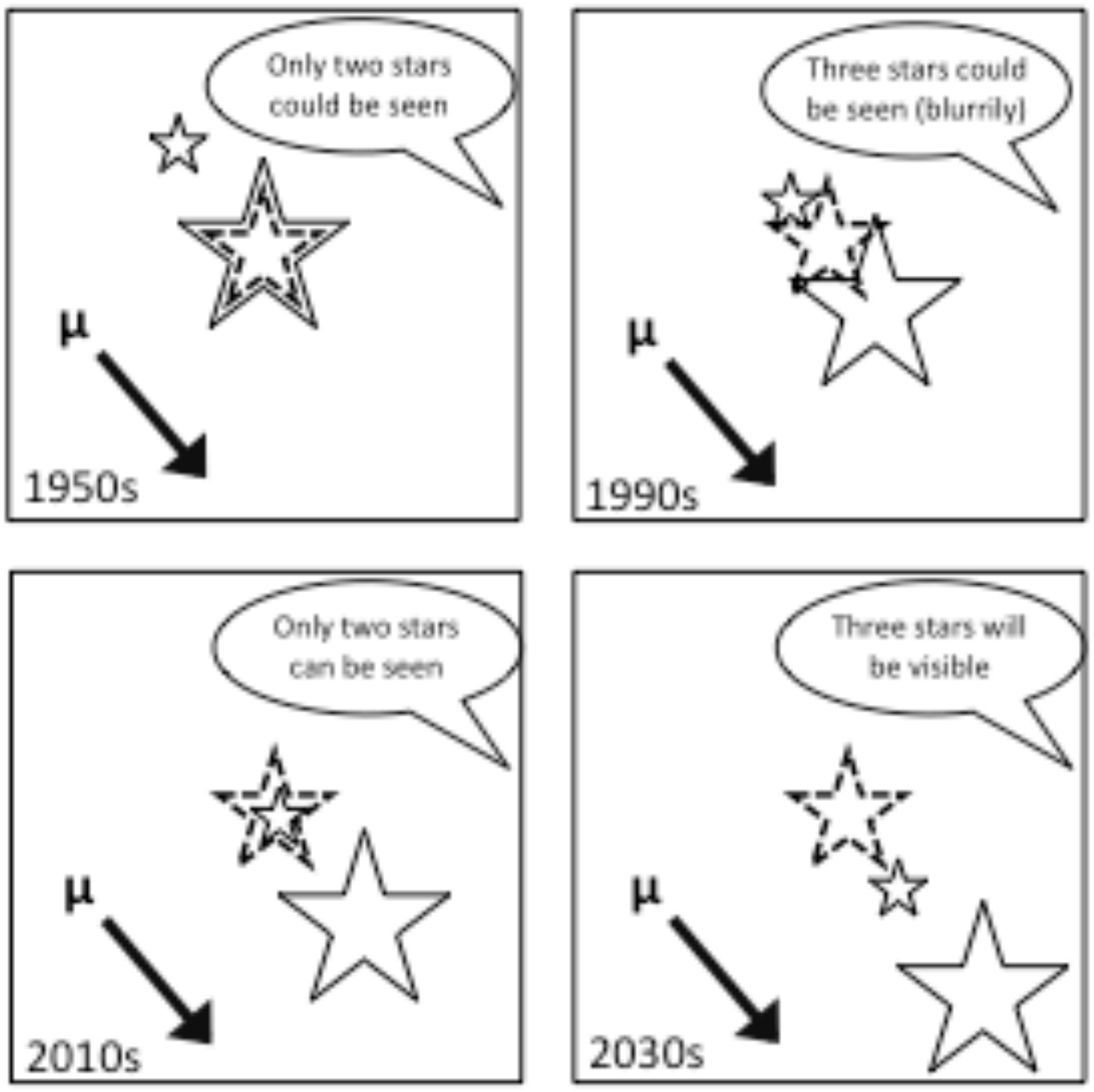

FIG. 2

Sketch explaining the apparent "triple" system WDS 07397+3328, formed by a physical binary of bright and faint stars (solid lines) moving to the southwest and a (fixed) background star of intermediate brightness (dashed lines). In the 1950s the bright primary overpowered the background star, while in the 2010s it is the background star that overpowers the faint secondary.

*Results*

For the 52 physical pairs (after discarding WDS 02457+4456 AB and WDS 10585−1046 in the previous section), we list their basic astrophysical parameters in Table IV. In particular, we provide:

*i.* *Spectral types of primaries and secondaries*. When available, they were taken from the most reliable sources, which were the Palomar/Michigan State University catalogue (PMSU)[32] and the preliminary results of the CARMENES science preparation[33–36]. Only in five cases were primaries earlier than K7 (*i.e.,* from F0 to K3), for which we took spectral types from the Simbad database[37–41]. There were also four systems containing white dwarfs, which are described below. Spectral typing was not available for four wide faint secondaries: LP 197−48 (WDS 02457+4456 B), PM I03510+1413 B (WDS 03510+1414 B), LP 256−44 (WDS 07397+3328 B), and G 264−18 B (WDS 22058+6539 B). For these, we estimated spectral types in the intervals m5: V to m7: V based on magnitude differences with respect to the primaries and a *J*-band absolute magnitude−spectral-type relation for M dwarfs[42] (we write 'm' instead of 'M' for spectral types derived from photometry). Of these, only PM I03510+1413 B had another photometric spectral-type estimation (m5 V[43], identical to ours). As described below, we also estimated photometric spectral types of 13 M dwarfs in very close pairs (not resolved in our images) from published magnitude differences and improved previous determinations of two resolved secondaries based on available data. Except for a few cases (*e.g.,* τ Cyg Ca and Cb), all our spectral-type estimations are new.

ii. *Heliocentric distances, d*. For 45 cases, the distances to the primaries were parallactic and compiled from the literature[22,44–46]. For the other seven cases, tabulated in parenthesis, we calculated spectro-photometric distances from a custom-made quadratic $M_J$–spectral-type relation. For deriving it, we compiled a ~1000-star sample of M dwarfs with *J*-band apparent magnitudes and parallactic distances, to be used for another programme (Cortés-Contreras, in prep.). As shown in Table V, calculated values reasonably match, and perhaps improve, previous spectro-photometric distance estimations[32,43]. All systems except six are located at less than 20 pc; seven systems are located at less than 10 pc.

iii. *Projected physical separations, s*. They are just the product $s = \rho\ d$, $\rho$ being the angular separation listed in Table I (in epoch interval 2012.6–2013.1).

iv. *Individual masses for both primary and secondary, $M_1$ and $M_2$*. Masses of M dwarfs were calculated with a spectral-type–mass relationship used internally by the CARMENES Consortium[4] and checked with previous estimations, when available[47]. Masses of primaries with earlier spectral types and of white dwarfs were obtained from the literature[41,48–51].

v. *Reduced orbital periods, P**. They were computed from the relationship $(M_1 + M_2)\ P^2 = a_2^3$ (in convenient units), where $a = a_1 + a_2$, $M_1\ a_1 = M_2\ a_2$ and the semi-major axis *a* was replaced by the projected physical separation *s*. Calculated reduced periods range from slightly over one century to several millenia. Reduced periods match actual ones to within a factor of three, depending on actual eccentricity[52].

vi. *Reduced binding energies*[13], $-U_g^*$. We used $U_g^* = -G\ M_1\ M_2\ s^{-1}$, where again the actual physical separation *r*, which can be approximated by *a* at low eccentricities, was replaced by *s*. In the case of wide multiple systems with very long orbital periods, of over one thousand years, for which it is very hard (if not impossible) to determine orbits and semi-major axes, $-U_g^*$ allows easy comparison of binding energies of systems published by different authors[53,54].

*M dwarf–white dwarf binaries*

Four of our pairs contain degenerate remnants: WDS 07307+4813 (with EGGR 52 AB; $5.9^{+2.7}_{-1.4}$ Gyr[50]), WDS 11080–0509 (with EGGR 76; $3.9^{+1.4}_{-1.3}$ Gyr[50]), WDS 13484+2337 (with EGGR 438; 4.72 Gyr[51]) and WDS 20568–0449 (with EGGR 202; 4.27 Gyr[51]). Interestingly, the system including EGGR 52 AB may be triple (see below). Although the white dwarfs are listed in Table I (and in WDS) as secondaries because they are fainter in the optical and near-infrared than the M dwarfs, they are more massive[41,50,51,55,56]. Furthermore, masses of the main-sequence stars that evolved into the white dwarfs were in all cases above one solar mass.

Actual ages of the four systems are longer than the cooling times of the remnants, which lie in the narrow interval between 3.9 and 5.9 Gyr[50,51]; once the stellar progenitor's main-sequence lifetime is added, actual system ages are considerably older (> 6−7 Gyr).

Given the relatively large projected physical separations of the four pairs, from 260 to 4800 AU, we can assume that the components have evolved as single stars (but see Morgan *et al.*[57]) and that there may exist a dynamical evolution associated with the remnant progenitor evolution (*i.e.,* an increase of the physical separation when the star quits the main sequence and loses mass).

*Hierarchical triple and quadruple systems*

As shown in Table VI, there are 16 stars in 14 systems that are close binaries unresolved in our IAC80 images. Three of them are double-lined spectroscopic binaries: WDS 08082+2106 Ba,Bb (BD+21 1764Ba,Bb[58]), WDS 08427+0935 Aa,Ab (BD+10 1857Aa,Ab[59]) and WDS 23573−1259 Ba,Bb (LP 704−014 AB[60]). The other 13 stars have been resolved with high-resolution imagers (*i.e., Hubble Space Telescope*, adaptive optics, lucky imaging, speckle). Fig. 3 illustrates the following discussion.

The WDS 21148+3803 system, composed of τ Cyg AB and τ Cyg Cab, is one of the only three quadruples in our sample. It was thought to be a quintuple because of a low-metallicity, low-mass, esdK-type star companion candidate at about 534 arcsec to the primary and out of the IAC80 field of view, namely LSR J2115+3804 (LEP 100 I[61]). A simple astrometric analysis discards it as a physical quintuple system because of the different proper motions (Cortés-Contreras, in prep.). Besides, Daley's companion candidate to τ Cyg AB, DAL 38 G[62], is also a background star according to PPMXL data.

The quadruple system WDS 08427+0935 is composed of a 21-d spectroscopic binary[59], which is the brightest component of a 62-yr astrometric double, which in turn has a wide, proper-motion, M-dwarf companion of lower mass imaged with TCP. As a result, this system has an interesting architecture of consecutive 'pairs' of stars separated by ~0.16, 18.9, and 1770 AU. Perhaps not by chance, both the WDS 21148+3803 and WDS 08427+0935 quadruples are, with over 2 $M_{\odot}$, the most massive of our systems.

The other quadruple system in our sample, of slightly over 1 $M_{\odot}$, is WDS 12576+3514. It consists of the variable star BF CVn (M1.5 V), a double companion at about 4 arcmin (M4.0 V+m4: V, for which $\Delta Ks$ has only been provided[63]) and a wide proper-motion companion of only about 0.08 $M_{\odot}$ located at 12.7 arcmin (approximately 14700 AU) to the primary and outside the TCP field of view, namely LP 268−004 (LEP 60[64,65]).

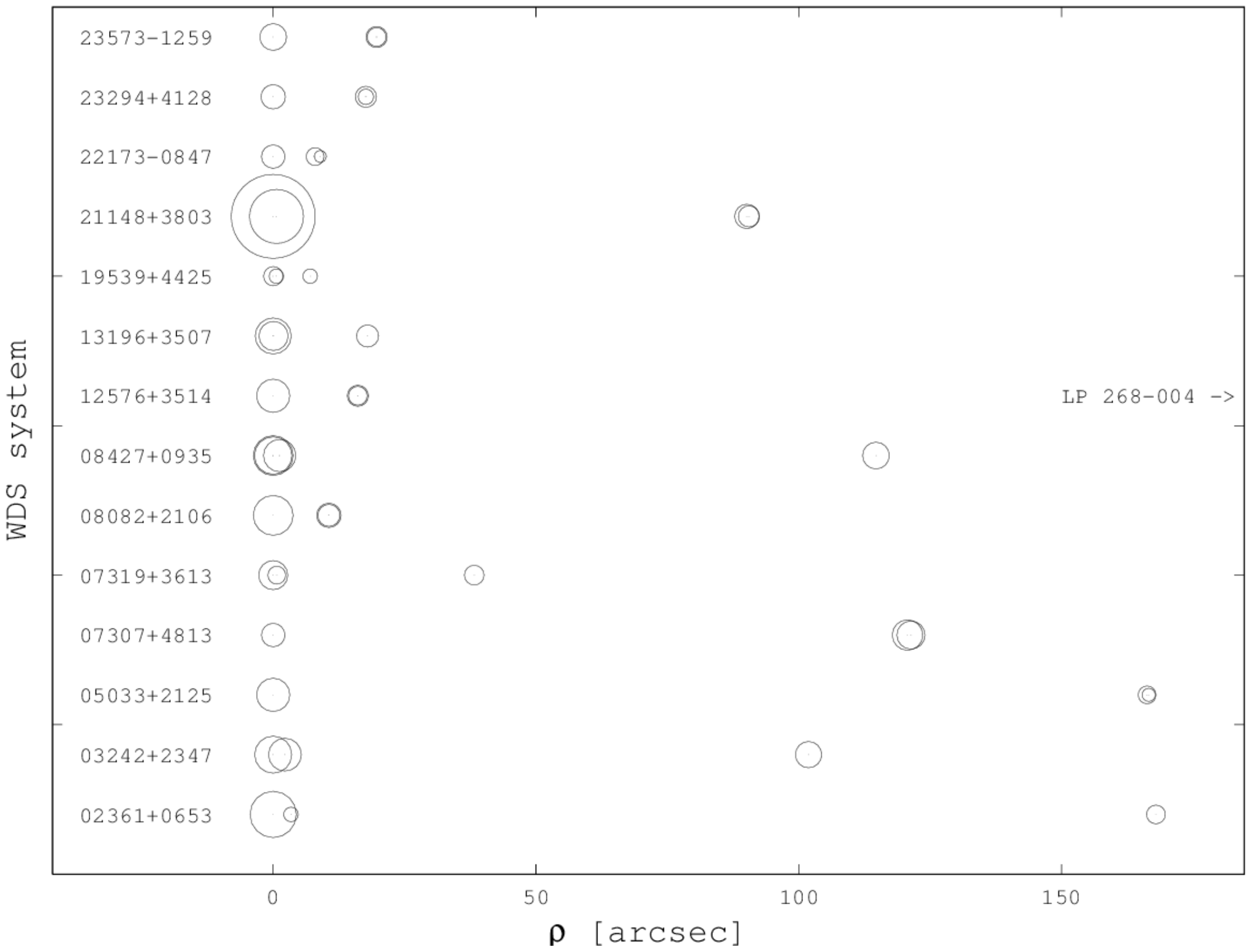

FIG. 3
Architecture of triple and quadruple star systems in our sample. Sizes of circles are approximately proportional to star mass. Primaries at located at $\rho$ = 0 arcsec. The quaternary of WDS 12576+3514 (LP 268−004) is out of the diagram at $\rho$ = 760.2 arcsec.

As previously noticed, since our initial sample of 106 observing fields was biased towards multiple systems, it would be incorrect to derive here a frequency of binaries. However, we can get some reliable statistics on the minimum frequency of higher-order multiples since there was no *a priori* bias towards them in our sample. There are at least 11 triples and three quadruples among our list of 52 physical systems studied (there can be still unknown close binaries or very wide proper-motion companions awaiting discovery), which results in minimum frequencies of triples and quadruples among multiple systems with M dwarf components of $21^{+11}_{-8}$% and $6^{+5}_{-4}$% (*i.e.,* roughly one out of five 'binaries' is actually a triple, and roughly one out of 17 'binaries' is actually a quadruple). The non-detection of quintuples (or higher order) points to a frequency of them among multiple systems with M-dwarf components of less than 1.9%. For comparison, see Tokovinin[48] for frequency and statistics of triple and quadruple systems among multiple systems at all stellar masses.

From the variety of system architectures in Table VI and Fig. 3, we do not see any relation between $\rho_{\rm wide}/\rho_{\rm close}$ ratios (from about 8 to over 500) and mass ratios, probably because of the small sample size if compared with previous comprehensive, dedicated works[48]. However, although tertiary companions tend to have masses comparable to the components of the inner binary, we find that the fraction of triple systems where the outer companion has the smallest mass is significantly larger than previously measured (about two thirds, in comparison with 46% measured by Tokovinin[48]). Besides, systems with large $\rho_{\rm wide}/\rho_{\rm close}$ ratios are expected to be in stable configurations; however, the compact systems with low $\rho_{\rm wide}/\rho_{\rm close}$ ratios, such as WDS 19539+4425 (GIC 159 + MCY 3) and WDS 22173−0847 (LDS 182 + BEU 22), may deserve further study of long-term dynamical stability, spin alignment, or possible orbital evolution. These studies may be also extended to the hierarchical quadruple system WDS 08427+0935 (LUY 6218 + ST 8).

*The most fragile systems*

We have identified five systems with reduced binding energies $-U_g^*$ equal to or below $10^{35}$ J. Two of them are the M-dwarf–white-dwarf pairs WDS 11080−0509 and WDS 13484+2337, whose actual binding energies depend closely on a better estimation of the white-dwarf masses. The other three fragile systems are composed of M dwarfs only: WDS 05033+2125 (which is actually a triple system), WDS 20555−1400, and WDS 21440+1705. The five of them have parallactic distance measurements and projected physical separations between 1000 and 4600 AU, approximately; there are other systems in our sample with similar projected physical separations but larger binding energies because of their larger (product of) mass. WDS 21440+1705 could be even more fragile than tabulated if the secondary were later than M4.5 V (see below). According to the Weinberg *et al.* [77] relationships, and with the parameters given Close *et al.*[53] and Dhital *et al.*[54], the five systems can survive in the Galactic disc for amounts of time comparable to the age of the Universe.

*Binaries with periods shorter than one millenium*

We looked for extra publicly-available data on the pairs with the shortest orbital periods. Our typical astrometry baseline coverage was of 60 yr (*e.g.,* since the USNO-A2 epoch of the first Palomar Observatory Sky Survey in the early 1950s to our recent epoch of IAC80 data). In some cases, when the Astrographic Catalogue AC2000.2[78] or very old WDS data were available, coverage was longer than one century. Within this time span, some systems, especially those with the shortest periods, displayed appreciable variations of position angle. In Table VII we list ten systems with the shortest reduced periods, of less than 1000 yr in all cases but one, and the largest $\theta$ variations ($\Delta\theta/360$), of up to over 50% of their orbits. Because of non-zero eccentricity, these fractions of orbit in position angle do not translate linearly into fractions of orbit in reduced period ($\Delta t/P^*$). In general, systems with larger/shorter fraction of orbit in $\theta$ than in $P^*$ were observed close to their periapses/apoapses. Only one ‘pair’, the triple system WDS 08082+2106 A–Bab (BD+21 1764), is close to its periapsis.

Very few of our systems have been the subject of orbital studies. For example, WDS 09144+5241 (STF 1321) shows a smooth change of position angle of 52.4 deg and in angular separation of 3.9 arcsec since the first astrometric epoch in 1821[79,80]. WDS 19464+3201 (KAM 3) also displays a large change of angular separation of 3.25 arcsec[81,82], which led some authors incorrectly to classify it as an unbound pair[83,84] (Cortés-Contreras et al., in prep.). Of the ten systems in Table VII, six displayed changed of the position angle larger than 12 deg, which is about 1/30 of a full orbit. In forthcoming papers we will investigate further all these systems that are interesting for the determination of dynamical masses of M dwarfs without the aid of high-resolution imagers.

*Final remarks*

In Table VIII, we summarize key comments on each star system investigated. This last table and the previous content is in the follow-on to the spirit of our series of papers in this pages, which is to shed light on how wide multiple systems with M-dwarf components could form and evolve by way of a simple but detailed examination of some selected systems.

*Acknowledgements*

We thank Bill Hartkopf for his enthusiastic referee report, Laura Toribio san Cipriano for her service-mode observations at IAC80, F. Javier Alonso-Floriano and Alexis Klutsch for their spectral-typing advice, and Brian Mason for his prompt responses to our numerous WDS data requests. This article is based on observations with the IAC80 telescope operated by the Instituto de Astrofísica de Canarias in the Spanish Observatorio del Teide. Financial support was provided by the Spanish MICINN under grants AYA2011-30147-C03-02 (UCM) and AYA2011-30147-C03-03 (CAB).

# Table I

*Astrometry of 54 star pairs investigated with IAC80*

| *WDS* | *Disc.* | *Name 1* | *Name 2* | *ρ [arcsec]* | *θ [deg]* | *Epoch* |
|---|---|---|---|---|---|---|
| 00137+8038 | LDS 1503 | G 242-048 | LP 012-304 | 13.41±0.02 | 125.68±0.02 | 2012.683 |
| 00164+1950 | LDS 863 | EZ Psc | LP 404-062 | 25.07±0.03 | 58.2±0.2 | 2012.683 |
| 00184+4401 | GRB 34 | GX And | GQ And | 34.5±0.4 | 65.1±0.4 | 2012.683 |
| 01119+0455 | GIC 20 | LHS 1212 | LHS 1213 | 63.60±0.02 | 145.69±0.02 | 2012.703 |
| 02361+0653 | PLQ 32 | HD 16160 AB | BX Cet | 164.6±0.4 | 109.1±0.4 | 2012.716 |
| 02457+4456 | LDS 5393 | G 078-004 | LP 197-048 | 17.90±0.10 | 65.9±0.3 | 2012.716 |
| | *GIC 34* | | *G 078-003* | *89.11±0.10* | *267.6±0.3* | *2012.716* |
| 02565+5526 | LDS 5401 | Ross 364 | Ross 365 | 16.9±0.4 | 20.5±0.4 | 2012.719 |
| 03242+2347 | LDS 884 | GJ 140 AB | GJ 140 C | 99.62±0.10 | 118.2±0.3 | 2012.714 |
| 03398+3328 | ES 327 | HD 278874 A | HD 278874 B | 15.56±0.40 | 293.6±0.4 | 2012.719 |
| 03510+1414 | JLM 1 | PM I03510+1413 A | PM I03510+1413 B | 28.74±0.02 | 319.01±0.02 | 2012.719 |
| 05033+2125 | LDS 6160 | HD 285190 A | HD 285190 BC | 166.30±0.10 | 240.9±0.3 | 2012.719 |
| 05342+1019 | LDS 6189 | Ross 45 A | Ross 45 B | 5.05±0.02 | 188.17±0.02 | 2012.809 |
| 05599+5834 | GIC 61 | EG Cam | G 192-012 | 161.04±0.10 | 119.4±0.3 | 2012.809 |
| 06007+6809 | LDS 1201 | LP 057-041 | LP 057-040 | 56.07±0.02 | 196.53±0.02 | 2012.812 |
| 06423+0334 | GIC 65 | G 108-021 | G 108-022 | 49.99±0.02 | 39.79±0.02 | 2012.812 |
| 07307+4813 | GIC 75 | GJ 275.2 A | EGGR 52 AB | 102.70±0.02 | 153.67±0.02 | 2012.812 |
| 07319+3613 | LDS 6206 | VV Lyn AB | BL Lyn | 37.57±0.10 | 352.2±0.3 | 2012.812 |
| 07397+3328 | LDS 3755 | G 090-016 | LP 256-044 | (13.66±0.02) | (48.84±0.02) | 2012.812 |
| 08082+2106 | COU 91 | BD+21 1764 A | BD+21 1764 Bab | 10.6±0.4 | 144.0±0.4 | 2012.866 |
| 08313-0601 | LDS 221 | LP 665-021 | LP 665-022 | 84.35±0.02 | 97.06±0.02 | 2012.867 |
| 08427+0935 | LUY 6218 | BD+10 1857 AabB | BD+10 1857 C | 114.7±0.4 | 96.2±0.4 | 2012.927 |
| 08526+2820 | LDS 6219 | ρ Cnc A | ρ Cnc B | 85.1±0.4 | 127.6±0.4 | 2012.927 |
| 09008+0516 | OSV 2 | Ross 686 | Ross 687 | 29.45±0.02 | 115.52±0.02 | 2012.927 |
| 09144+5241 | STF 1321 | HD 79210 | HD 79211 | 17.2±0.4 | 96.4±0.4 | 2012.954 |
| 09288-0722 | GIC 87 | Ross 439 A | Ross 439 B | 35.59±0.03 | 83.4±0.2 | 2012.954 |
| 09427+7004 | OSV 3 | GJ 360 | GJ 362 | 89.0±0.4 | 76.6±0.4 | 2012.954 |
| 10261+5029 | LDS 1241 | LP 127-371 | LP 127-372 | 14.37±0.02 | 25.36±0.02 | 2012.954 |
| 10585-1046 | *LDS 4041* | *BD-10 3166* | *LP 731-076* | *21.6±0.4* | *212.7±0.4* | *2012.957* |
| 11055+4332 | VBS 18 | BD+44 2051A | WX UMa | 31.6±0.4 | 124.5±0.4 | 2012.957 |
| 11080-0509 | LDS 852 | GJ 1142 A | EGGR 76 | 278.80±0.02 | 339.0±0.2 | 2012.957 |
| 12123+5429 | VYS 5 | BD+55 1519A | BD+55 1519B | 14.8±0.4 | 11.8±0.4 | 2012.957 |
| 12576+3514 | LDS 5764 | BF CVn | BD+36 2322Bab | 16.1±0.4 | 226.0±0.4 | 2012.957 |
| 13196+3507 | HJ 529 | BD+35 2436Aab | BD+35 2436B | 17.9±0.4 | 132.2±0.4 | 2012.957 |
| 13484+2337 | LDS 4410 | GJ 1179 A | EGGR 438 | 187.49±0.02 | 229.61±0.02 | 2013.026 |
| 18180+3846 | GIC 151 | LHS 462 | LHS 461 | 9.97±0.03 | 277.2±0.2 | 2012.768 |
| 19072+2053 | LDS 1017 | HD 349726 | Ross 731 | 114.5±0.4 | 289.9±0.4 | 2012.755 |
| 19147+1918 | LDS 1020 | Ross 733 | Ross 734 | 40.80±0.02 | 178.71±0.02 | 2012.755 |
| 19169+0510 | LDS 6334 | V1428 Aql | V1298 Aql (vB 10) | 75.8±0.4 | 152.1±0.4 | 2012.755 |
| 19464+3201 | KAM 3 | BD+31 3767A | BD+31 3767B | 5.5±0.4 | 134.7±0.4 | 2012.755 |
| 19510+1025 | J 124 | o Aql A | o Aql B | 21.3±0.4 | 218.2±0.4 | 2012.755 |
| 19539+4425 | GIC 159 | V1581 Cyg AB | GJ 1245 C | 6.45±0.02 | 70.42±0.02 | 2012.755 |
| 19566+5910 | GIC 161 | BD+58 2015A | BD+58 2015B | 72.89±0.10 | 253.2±0.3 | 2012.757 |
| 20446+0854 | LDS 1046 | LP 576-040 | LP 576-039 'AB' | 15.08±0.03 | 344.1±0.2 | 2012.757 |
| 20555-1400 | LDS 6418 | GJ 810 A | GJ 810 B | 107.19±0.02 | 184.61±0.02 | 2012.757 |
| 20568-0449 | LDS 6420 | FR Aqr | EGGR 202 (vB 11) | 14.94±0.03 | 309.8±0.2 | 2012.757 |
| 21011+3315 | LDS 1049 | LP 340-547 | LP 340-548 | 56.86±0.10 | 94.4±0.3 | 2012.741 |
| 21148+3803 | AGC 13 AF | τ Cyg AB | τ Cyg Cab | 89.5±0.4 | 184.2±0.4 | 2012.741 |
| 21161+2951 | LDS 1053 | Ross 776 | Ross 826 | 26.02±0.02 | 258.84±0.02 | 2012.741 |
| 21440+1705 | LDS 6358 | G 126-031 | G 126-030 | 64.09±0.02 | 346.01±0.02 | 2012.749 |
| 22058+6539 | NI 44 | G 264-018 A | G 264-018 B | 6.66±0.03 | 136.5±0.2 | 2012.749 |
| 22173-0847 | LDS 782 | FG Aqr A | Wolf 1561 BC | 7.98±0.02 | 214.40±0.02 | 2012.757 |
| 23294+4128 | GIC 193 | G 190-028 | G 190-027 AB | 17.65±0.02 | 213.70±0.02 | 2012.757 |
| 23573-1259 | LDS 830 | LP 704-015 | LP 704-014 AB | 19.69±0.02 | 294.44±0.02 | 2012.757 |

Table II

*Virtual-observatory astrometric follow-up of nine pairs*

| *WDS* | *Epoch* | *ρ [arcsec]* | *θ [deg]* | *Origin* |
|---|---|---|---|---|
| 02457+4456 | 1951.971 | 80.74 | 269.50 | POSS-I Red |
| | 1983.021 | 84.79 | 268.66 | GSC 1.3 |
| | 1989.748 | 86.15 | 268.54 | POSS-II Red |
| | 1998.868 | 87.14 | 268.27 | 2MASS |
| | 2003.084 | 87.74 | 268.03 | SDSS-DR9 |
| | 2004.010 | 86.57 | 268.20 | POSS-II Blue |
| | 2004.067 | 87.74 | 268.12 | CMC14 |
| | 2005.535 | 86.69 | 268.26 | POSS-II Infrared |
| | 2010.096 | 88.71 | 267.95 | *WISE* |
| | 2012.716 | 89.11 | 267.60 | *This work* [TCP] |
| | *Average* | ... | ... | |
| 03398+3328 | 1955.807 | 15.20 | 293.47 | POSS-I Red |
| | 1988.716 | 15.75 | 293.87 | POSS-II Red |
| | 1989.669 | 15.91 | 295.82 | POSS-II Blue |
| | 1990.882 | 15.14 | 293.58 | POSS-II Infrared |
| | 1998.758 | 15.43 | 294.46 | 2MASS |
| | 2010.118 | 15.41 | 294.28 | *WISE* |
| | 2012.719 | 15.56 | 293.57 | *This work* [TCP] |
| | *Average* | 15.48±0.28 | 294.15±0.83 | *This work* |
| 07397+3328 | 1955.117 | 13.03 | 56.30 | POSS-I Red |
| | 1990.079 | 12.67 | 59.11 | POSS-II Red |
| | 1998.170 | 12.78 | 57.42 | 2MASS |
| | 1998.988 | 12.64 | 56.49 | POSS-II Infrared |
| | 2007.139 | 13.07 | 58.88 | SDSS-DR9 |
| | 2012.812 | (13.66) | (48.84) | *This work* [TCP] |
| | *Average* | 12.84±0.20 | 57.6±1.3 | *This work* |
| 09288−0722 | 1953.929 | 36.04 | 84.54 | POSS-I Red |
| | 1984.177 | 35.79 | 83.76 | GSC 2.2 |
| | 1986.000 | 35.78 | 83.68 | UKST Infrared |
| | 1991.268 | 35.51 | 83.44 | UKST Red |
| | 1999.046 | 35.83 | 83.96 | 2MASS |
| | 1999.771 | 35.60 | 83.94 | UKST Blue |
| | 2004.067 | 35.69 | 83.62 | CMC14 |
| | 2010.355 | 35.62 | 83.59 | *WISE* |
| | 2012.957 | 35.59 | 83.35 | *This work* [TCP] |
| | *Average* | 35.73±0.17 | 83.8±0.3 | *This work* |
| 10585−1046 | 1954.246 | 18.01 | 219.67 | POSS-I Red |
| | 1986.285 | 19.66 | 216.15 | GSC 1.3 |
| | 1992.037 | 20.44 | 215.43 | UKST Red |
| | 1995.218 | 20.63 | 215.16 | UKST Infrared |
| | 1993.124 | 20.75 | 214.47 | 2MASS |
| | 2004.059 | 20.99 | 213.87 | CMC14 |
| | 2010.423 | 21.46 | 213.67 | *WISE* |
| | 2012.957 | 21.61 | 212.74 | *This work* [TCP] |
| | *Average* | ... | ... | |
| 12123+5429 | 1955.284 | 14.32 | 9.92 | POSS-I Red |
| | 1984.163 | 14.40 | 8.94 | GSC 1.3 |
| | 1991.347 | 14.57 | 10.00 | POSS-II Red |
| | 1995.309 | 14.19 | 8.05 | POSS-II Blue |
| | 1998.167 | 14.61 | 10.47 | POSS-II Infrared |
| | 1999.339 | 14.66 | 9.50 | 2MASS |
| | 2001.962 | 14.62 | 9.58 | SDSS-DR9 |
| | 2010.371 | 14.63 | 9.69 | *WISE* |
| | 2012.957 | 14.81 | 11.79 | *This work* [TCP] |
| | *Average* | 14.53±0.19 | 9.77±1.03 | *This work* |
| 20446+0854 | 1953.751 | 14.90 | 343.92 | POSS-I Red |
| | 1984.511 | 14.78 | 343.21 | GSC 1.3 |
| | 1987.492 | 14.88 | 345.04 | POSS-II Red |
| | 1990.624 | 14.91 | 343.79 | POSS-II Blue |
| | 1994.441 | 15.00 | 345.16 | POSS-II Infrared |
| | 2000.349 | 15.13 | 344.06 | 2MASS |
| | 2000.738 | 15.07 | 344.02 | SDSS-DR9 |
| | 2002.364 | 15.10 | 344.21 | CMC14 |
| | 2010.337 | 15.00 | 344.51 | WISE |
| | 2012.784 | 15.08 | 344.11 | *This work* [TCP] |
| | *Average* | 14.98±0.11 | 344.21±0.58 | *This work* |
| 22058+6539 | 1954.594 | 6.97 | 135.55 | POSS-I Red |
| | 1991.697 | 6.48 | 134.43 | POSS-II Red |
| | 1993.554 | 6.59 | 133.88 | POSS-II Blue |
| | 1994.586 | ~6.7 | ~135 | POSS-II Infrared |
| | 1999.744 | 6.76 | 135.70 | 2MASS |
| | 2010.025 | 6.65 | 137.11 | WISE |
| | 2012.749 | 6.66 | 136.51 | *This work* [TCP] |
| | *Average* | 6.69±0.17 | 135.53±1.22 | *This work* |
| 23294+4128 | 1952.631 | 17.69 | 211.02 | POSS-I Red |
| | 1984.645 | 17.52 | 211.09 | GSC 1.3 |
| | 1987.793 | 17.69 | 214.98 | POSS-II Blue |
| | 1989.751 | 17.60 | 213.59 | POSS-II Red |
| | 1995.624 | 17.72 | 213.97 | POSS-II Infrared |
| | 1999.755 | 17.67 | 213.89 | 2MASS |
| | 2002.597 | 17.68 | 213.84 | CMC14 |
| | 2010.497 | 17.65 | 213.96 | *WISE* |
| | 2012.757 | 17.65 | 213.70 | *This work* [TCP] |
| | *Average* | 17.65±0.06 | 213.7±1.1 | *This work* |

Table III

*Catalogued and measured proper motions of physical pairs VYS 5 and GIC 193*

| *WDS* | *A* | | *B* | | *Ref.* |
|---|---|---|---|---|---|
| | $\mu_\alpha cos\delta$ *[mas yr⁻¹]* | $\mu_\delta$ *[mas yr⁻¹]* | $\mu_\alpha cos\delta$ *[mas yr⁻¹]* | $\mu_\delta$ *[mas yr⁻¹]* | |
| 12123+5429 (VYS 5) | +250 | +100 | +250 | +100 | Giclas[29] |
| | +232 | +90 | ... | ... | USNO-B1[31] |
| | +233 | +91 | +182 | +37 | LSPM[30] |
| | +231.5±1.3 | +89.9±1.3 | ... | ... | HIP[22] |
| | +245.3±3.9 | +85.9±2.1 | +245.6±4.2 | +92.0±2.1 | *This work* |
| 23294+4128 (GIC 193) | +460 | −50 | +460 | −50 | Giclas[29] |
| | ... | ... | +220 | −82 | USNO-B1[31] |
| | ... | ... | +224.8 | −83.5 | PPMXL[23] |
| | +415 | −41 | +415 | −41 | LSPM[30] |
| | +412.8±3.3 | −53.5±5.1 | +400.0±4.2 | −45.3±2.4 | *This work* |

## Table IV
*Basic astrophysical data of 52 physical pairs*

| *WDS* | *Sp. Type 1* | *Sp. Type 2* | *d [pc]* | *s [AU]* | *$M_1$ [$M_\odot$]* | *$M_2$ [$M_\odot$]* | *P* [$10^3$ yr]* | *$-U^*_g$ [$10^{33}$ J]* |
|---|---|---|---|---|---|---|---|---|
| 00137+8038 | M1.5V | M5.0V | 19.6±0.7 | 263±9 | 0.46±0.05 | 0.18±0.04 | 3.2±0.2 | 560±140 |
| 00164+1950 | M4.0V | M4.0V | 16±4 | 410±90 | 0.25±0.03 | 0.25±0.03 | 4.1±1.3 | 270±70 |
| 00184+4401 | M1.0V | M3.5V | 3.587±0.010 | 123.8±1.5 | 0.50±0.04 | 0.27±0.05 | 0.82±0.04 | 1900±400 |
| 01119+0455 | M3.0V | M3.5V | 15.9±0.5 | 1010±30 | 0.32±0.05 | 0.27±0.05 | 16.7±1.3 | 150±40 |
| 02361+0653 | K3V+M7.0V | M4.0V | 7.18±0.02 | 1182±4 | 0.85±0.03 | 0.25±0.03 | 26.3±0.6 | 320±40 |
| 02457+4456 | M0.5V | m6:V | 23.1±1.2 | 410±20 | 0.52±0.02 | ~0.13±0.03 | ~7.5±0.6 | ~290±70 |
| 02565+5526 | M1.0V | M3.0V | 19.5±1.8 | 330±30 | 0.50±0.04 | 0.32±0.05 | 3.1±0.5 | 860±170 |
| 03242+2347 | M0.0V+m1:V | M2.0V | 19.5±1.0 | 1940±100 | 1.04±0.06 | 0.41±0.05 | 43±4 | 390±60 |
| 03398+3328 | K5V | M3.0V | 43±3 | 670±50 | 1.30±0.10 | 0.32±0.05 | 9.9±1.2 | 1100±200 |
| 03510+1414 | M4.5V | m5:V | (14.4±1.0) | 410±30 | 0.22±0.04 | ~0.18±0.04 | ~5.4±0.7 | ~170±50 |
| 05033+2125 | M1.5V | M5.0V+m6:V | 27±6 | 4600±1000 | 0.46±0.05 | ~0.31±0.07 | ~160±60 | ~55±19 |
| 05342+1019 | M3.0V | M4.5V | (17.5±0.7) | 89±3 | 0.32±0.05 | 0.22±0.04 | 0.52±0.04 | 1400±300 |
| 05599+5834 | M0.5V | M4.0V | 13.5±0.3 | 2180±50 | 0.52±0.02 | 0.25±0.03 | 64±3 | 105±13 |
| 06007+6809 | M3.5V | M4.0V | 20.1±0.7 | 1130±40 | 0.27±0.05 | 0.25±0.03 | 19.6±1.5 | 110±20 |
| 06423+0334 | M3.5V | M4.0V | 12.8±0.5 | 640±20 | 0.27±0.05 | 0.25±0.03 | 8.4±0.6 | 190±40 |
| 07307+4813 | M4.0V | DA10+da10 | 11.10±0.10 | 1140±10 | 0.25±0.03 | ~0.90±0.04 | ~24.9±0.6 | ~450±40 |
| 07319+3613 | M2.5V+m5:V | M3.5V | 11.9±0.5 | 446±18 | ~0.55±0.09 | 0.27±0.05 | ~5.7±0.5 | ~590±150 |
| 07397+3328 | M2.0V | m6:V | 36±4 | 490±50 | 0.41±0.05 | ~0.13±0.03 | ~9.7±1.7 | ~190±50 |
| 08082+2106 | K7V | M3.0V+m3:V | 16.6±0.5 | 176±9 | 0.60±0.10 | ~0.64±0.10 | ~0.70±0.07 | ~3900±900 |
| 08313-0601 | M1.5V | M3.0V | 26.4±1.9 | 2220±160 | 0.46±0.05 | 0.32±0.05 | 54±6 | 120±20 |
| 08427+0935 | K7V+k7:V+M0.0V | M2.5V | 15.4±0.6 | 1770±60 | 1.74±0.22 | 0.37±0.05 | 38±3 | 640±120 |
| 08526+2820 | G8V | M4.5V | 12.34±0.11 | 1051±11 | 0.90±0.10 | 0.22±0.04 | 23.2±1.2 | 330±70 |
| 09008+0516 | M3.0V | M3.5V | 21±3 | 620±100 | 0.32±0.05 | 0.27±0.05 | 8±2 | 250±70 |
| 09144+5241 | M0.0V | M0.0V | 5.8±0.2 | 100±4 | 0.54±0.02 | 0.54±0.02 | 0.34±0.02 | 5200±400 |
| 09288-0722 | M2.5V | M4.5V | 16.2±1.0 | 580±40 | 0.37±0.05 | 0.22±0.04 | 9.0±1.0 | 250±60 |
| 09427+7004 | M2.0V | M3.0V | 12.3±0.3 | 1090±20 | 0.41±0.05 | 0.32±0.05 | 17.8±1.0 | 210±40 |
| 10261+5029 | M4.0V | M4.0V | (16.6±1.0) | 239±14 | 0.25±0.03 | 0.25±0.03 | 1.85±0.18 | 460±80 |
| 11055+4332 | M1.0V | M5.0V | 4.85±0.02 | 154±2 | 0.50±0.04 | 0.18±0.04 | 1.45±0.07 | 1000±200 |
| 11080-0509 | M3.0V | DA3.1 | 17±4 | 4800±1000 | 0.32±0.05 | 0.56±0.03 | 180±60 | 66±18 |
| 12123+5429 | M0.0V | M3.0V | 15.5±0.3 | 230±8 | 0.54±0.02 | 0.32±0.05 | 1.87±0.11 | 1300±200 |
| 12576+3514 | M1.5V | M4.0V+m4:V | 19.3±1.1 | 310±19 | 0.46±0.05 | ~0.50±0.04 | ~1.84±0.18 | ~1300±200 |
| 13196+3507 | M0.5V+m0.5:V | M3.0V | 13.3±0.3 | 237±7 | ~1.04±0.04 | 0.32±0.05 | ~2.10±0.11 | ~2500±400 |
| 13484+2337 | M5.5V | DA10 | 12.1±0.3 | 2270±60 | 0.16±0.03 | 0.45±0.02 | 88±4 | 58±11 |
| 18180+3846 | M3.0V | M4.0V | 10.7±0.4 | 107±4 | 0.32±0.05 | 0.25±0.03 | 0.61±0.05 | 1300±300 |
| 19072+2053 | M2.0V | M2.0V | 8.51±0.16 | 974±19 | 0.41±0.05 | 0.41±0.05 | 11.9±0.6 | 300±50 |
| 19147+1918 | M3.5:V | M3.5V | 19.1±1.1 | 780±40 | 0.27±0.05 | 0.27±0.05 | 10.4±1.1 | 170±40 |
| 19169+0510 | M2.5V | M8.0V | 5.87±0.03 | 445±3 | 0.37±0.05 | 0.10±0.01 | 9.6±0.5 | 150±20 |
| 19464+3201 | M0.5V | M2.0V | 13.6±0.3 | 75±6 | 0.52±0.02 | 0.41±0.05 | 0.28±0.03 | 5000±700 |
| 19510+1025 | F8V | M3.5V | 19.19±0.11 | 408±8 | 1.32±0.15 | 0.27±0.05 | 5.0±0.3 | 1500±300 |
| 19539+4425 | M5.5V+m7:V | M5.5V | 4.56±0.03 | 29.4±0.2 | ~0.27±0.04 | 0.16±0.03 | ~0.121±0.007 | ~2600±600 |
| 19566+5910 | K7V | M3.5V | 31.7±1.8 | 2310±130 | 0.60±0.10 | 0.27±0.05 | 68±7 | 120±30 |
| 20446+0854 | M1.5V | M3.5V+m3.5:V | (21.8±0.4) | 330±6 | 0.46±0.05 | 0.54±0.10 | 1.86±0.12 | 1300±300 |
| 20555-1400 | M4.0V | M5.0V | 13±3 | 1400±300 | 0.25±0.03 | 0.18±0.04 | 37±13 | 56±19 |
| 20568-0449 | M4.0V | DC10 | 17.7±1.2 | 264±18 | 0.25±0.03 | 0.37±0.06 | 2.5±0.3 | 620±130 |
| 21011+3315 | M3.0V | M3.5V | (16.6±0.7) | 940±40 | 0.32±0.05 | 0.27±0.05 | 15.0±1.3 | 160±40 |
| 21148+3803 | F0IV+G1V | M2.5V+m4:V | 20.34±0.16 | 1820±16 | 2.5±0.2 | ~0.62±0.06 | ~31.5±1.4 | ~1500±200 |
| 21161+2951 | M3.5V | *m4.5:V* | (13.9±0.7) | 362±18 | 0.27±0.05 | 0.22±0.04 | 4.0±0.4 | 290±80 |
| 21440+1705 | M4.0V | M4.5V | 15.3±0.6 | 980±40 | 0.25±0.03 | 0.22±0.04 | 17.4±1.4 | 100±20 |
| 22058+6539 | M3.5V | m6:V | (13.7±0.7) | 92±4 | 0.27±0.05 | ~0.13±0.03 | ~0.77±0.08 | ~700±300 |
| 22173-0847 | M4.0V | M5.0V+m7:V | 10.0±0.3 | 80±2 | 0.25±0.03 | ~0.29±0.05 | ~0.31±0.02 | ~1600±300 |
| 23294+4128 | M3.5V | M4.0V+m5:V | 14.9±0.5 | 263±9 | 0.27±0.05 | 0.43±0.07 | 1.22±0.10 | 780±190 |
| 23573-1259 | M3.0V | M4.0V+m5:V | (18.1±0.7) | 357±15 | 0.32±0.05 | ~0.43±0.07 | ~2.17±0.18 | ~680±160 |

Table V

*Spectrophotometric distances comparison*

| *WDS* | *d [pc] (Other authors)* | | *d [pc] (This work)* | |
|---|---|---|---|---|
| | *A* | *B* | *A* | *B* |
| 03510+1414 | 16.3[43] | 14.1[43] | 14.4±1.0 | ... |
| 05342+1019 | 10.9±3.3[32] | 28.0±8.4[32] | 17.5±0.7 | 12.9±0.7 |
| 10261+5029 | 18.0±5.4[32] | 19.2±5.8[32] | 16.6±1.0 | 17.7±1.0 |
| 20446+0854 | 17.9±5.4[32] | 12.0±3.6[32] | 21.8±0.4 | 15.0±0.7 (single)<br>21.2±1.0 (double) |
| 21011+3315 | 15.8±4.7[32] | 19.4±5.8[32] | 16.6±0.7 | 17.4±0.8 |
| 21161+2951 | 12.6±3.8[32] | 18.6±5.4[32] | 13.9±0.7 | *13.5±0.9* |
| 22058+6539 | 24.0±7.2[32] | ... | 13.7±0.7 | ... |
| 23573−1259 | 20.0±5.8[32] | 12.9±3.9[32] | 18.1±0.7 | 15.6±0.9 (single)<br>22.0±1.3 (double) |

Table VI

*Unresolved binaries in hierarchical triple (and quadruple) systems*

| *WDS* | *Disc.* | *Binary* | $\rho_{\rm close}$ *[arcsec]* | $\rho_{\rm wide}/\rho_{\rm close}$ | *Δmag (band) [mag]* |
|---|---|---|---|---|---|
| 02361+0653 | GKI 1 | HD 16160 AB | 3.394[66] | 62.6–48.5 | 10.84(*V*)[66] |
| | | | 2.63[67] | | 5.29(*Ks*)[67] |
| 03242+2347 | WOR 4 | GJ 140 AB | 2.247[22] | 44.4 | 1.27(*Hp*)[68] |
| 05033+2125 | LAW 13 | HD 285190 BC | 0.310[69] | 536 | 1.3(*i'*)[69] |
| | | | | | 0.8(*z'*)[69] |
| 07307+4813 | WNO 49 | EGGR 52 AB | 0.656[70] | 184 | 0.0(*V*)[70] |
| 07319+3613 | BEU 11 | VV Lyn | 0.684[71] | 54.9 | 2.02(*K*)[71] |
| 08082+2106 | ... | BD+21 1764 Bab[58] | <0.66[58] | >16 | ... |
| 08427+0935 | ST 8 | BD+10 1857 AabB | 1.236[72] | 92.8 | 3.9(800nm)[72] |
| | ... | BD+10 1857 Aa,Ab[59] | ... | ... | ... |
| 12576+3514 | ... | BD+36 2322 Bab | ... | ... | 0.061(*Ks*)[63] |
| 13196+3507 | BAG 11 | BD+35 2436 Aab | 0.065[73] | 275 | 0.0: (600nm)[73] |
| 19539+4425 | MCY 3 | V1581 Cyg AB | 0.594[74] | 10.9 | 1.29(F180M)[74] |
| | | | | | 1.18(F207M)[74] |
| | | | | | 1.08(F222M)[74] |
| 20446+0854 | ... | LP 576–039 'AB' | ... | ... | ... |
| 21148+3803 | AGC 13 AB | τ Cyg AB | 0.637[22] | 140 | 2.89(*Hp*)[68] |
| | JOD 20 | τ Cyg Cab | 0.40[75] | 224 | 1.55(*I*)[75] |
| 22173-0847 | BEU 22 | Wolf 1561 AB | 0.978[71] | 8.16 | 3.15(*z'*)[76] |
| | | | | | 1.18(*K*)[71] |
| 23294+4128 | ... | G 190–027 AB | ... | ... | 0.27(*H*)[63] |
| 23573–1259 | ... | LP 704-014 AB[60] | ... | ... | ... |

Table VII

*Physical systems with measured variations of position angle*

| *WDS* | *Disc.* | *Δθ* *[deg]* | *Δθ/360* | *Δt* *[yr]* | *Δt/P** | *Apsis* |
|---|---|---|---|---|---|---|
| 00184+4401 | GRB 34 | 25 | 0.07 | 150 | 0.19 | Apo. |
| 05342+1014 | LDS 6189 | 2.8 | 0.008 | 53 | 0.06 | ... |
| 08082+2106 A-Bab | COU 91 | 180 | 0.5 | 120 | 0.2 | Peri. |
| 09144+5241 | STF 1321 | 52 | 0.15 | 200 | 0.6 | Apo. |
| 11055+4332 | VBS 18 | 25 | 0.07 | 60 | 0.06 | ... |
| 18180+1846 | GIC 151 | 13 | 0.04 | 52 | 0.06 | ... |
| 19464+3201 | KAM 3 | 7.4 | 0.03 | 76 | 0.3 | Apo. |
| 19539+4425 AB-C | GIC 159 | 36 | 0.10 | 60 | 0.3 | Apo. |
| 22058+6539 | NI 44 | 0.76 | 0.002 | 13 | 0.02 | Apo. |
| 22173−0847 A-BC | LDS 782 | 1.9 | 0.005 | 63 | 0.2 | Apo. |

## Table VIII
*Summary of remarks on the 52 physical systems*

| *WDS* | *Remarks* |
|---|---|
| 00137+8038 | (LDS 1503) ... |
| 00164+1950 | (LDS 863) During the late 1990s and early 2000s, the primary of the binary system passed, from the observer's point of view, at only about 5 arcsec, by a Sun-like background star of null proper motion and similar brightness. This fact led Simbad and some authors to mix up the coordinates of the two stars. In 2014 and onwards, the primary of the true binary system is the southern star of the visual trio. |
| 00184+4401 | (GRB 34) Pair with period shorter than one millenium and appreciable orbital variation. |
| 01119+0455 | (GIC 20) ... |
| 02361+0653 | (PLQ 32 + GKI 1) Triple system (primary is a close double). |
| 02457+4456 | (LDS 5393) False triple, actual double system with astrometric follow-up; true secondary with estimated spectral type at m6:V. |
| 02565+5526 | (LDS 5401) ... |
| 03242+2347 | (LDS 884 + WOR 4) Triple system (primary is an active, close double). |
| 03398+3328 | (ES 327) Pair made of two Henry Draper stars of relatively low proper motion that needed astrometric follow-up. Several different spectral types have been proposed for the primary in the interval from K2IV to K5V[37,85,86]; its *Hipparcos* distance (43±3 pc) is significantly larger than the spectro-photometric distance to the secondary provided by Riaz *et al.*[87] (29 pc) or estimated by us with our custom-made SpT-$M_J$ relation (21 pc). We may need the ESA *Gaia* space mission and/or a detailed spectroscopic analysis to understand what is wrong with this system (age, metallicity, incorrect *Hipparcos* parallax of the primary, wrong PMSU and Riaz *et al.* spectral types, or unresolved multiplicity of the secondary). |
| 03510+1414 | (JLM 1) Poorly-known pair with spectro-photometric distance only[43,88]; secondary with estimated spectral type at m5:V. We propose for the first time that the primary, the secondary, or both are associated to the X-ray source 1RXS J035101.5+141404. |
| 05033+2125 | (LDS 6160 + LAW 13) Fragile, triple system (secondary is a close double). |
| 05342+1019 | (LDS 6189) Pair with period shorter than one millenium, spectro-photometric distance, and appreciable orbital variation. |
| 05599+5834 | (GIC 61) ... |
| 06007+6809 | (LDS 1201) ... |
| 06423+0334 | (GIC 65) ... |
| 07307+4813 | (GIC 75 + WNO 49) Triple system with a double white dwarf (EGGR 52 AB). The primary might in turn be a very close binary separated by 0.054 arcsec, which would make the system quadruple[70]. |
| 07319+3613 | (LDS 6206 + BEU 11) Triple system (primary is a close double). |
| 07397+3328 | (LDS 3755) Pair with astrometric follow-up; sketched in Fig. 2, showing background star; true secondary with estimated spectral type at m6:V. |
| 08082+2106 | (COU 91) Triple system with period shorter than one millenium and appreciable orbital variation. Secondary is a spectroscopic binary of unknown parameters[58]. |
| 08313-0601 | (LDS 221) TCP image in Fig. 1, left panel. |
| 08427+0935 | (LUY 6218 + ST 8) Quadruple system in a hierarchical arrangement with three orbital periods of about 0.057, 62, and 23200 yr containing a spectroscopic binary, and close 'pair' and a wide companion. |
| 08526+2820 | (LDS 6219) Pair with very bright primary of Bayer designation (ρ Cnc) that is also a bright multiplanet system host. |
| 09008+0516 | (OSV 2) ... |
| 09144+5241 | (STF 1321) Pair with period shorter than one millenium and appreciable orbital variation. |
| 09288-0722 | (GUC 87) Pair with astrometric follow-up. |
| 09427+7004 | (OSV 3) ... |
| 10261+5029 | (LDS 1241) Pair with spectro-photometric distance only. |
| 11055+4332 | (VBS 18) Pair with reduced period shorter than 1500 yr and appreciable orbital variation. |
| 11080-0509 | (LDS 852) Fragile system with a white dwarf (EGGR 76). |
| 12123+5429 | (VYS 5) Pair with discordant catalogued proper motions with astrometric follow-up. |
| 12576+3514 | (LDS 5764) Quadruple system made of an early M dwarf (BF CVn), a close-binary intermediate M dwarf, and a wide companion at the substellar boundary out of TCP field of view. The close pair has no WDS designation yet. |
| 13196+3507 | (HJ 529 + BAG 11) Triple system (primary is a close double) of relatively low proper motion. Although first measured in 1827[89], the first reliable astrometric measurement of the wide |

pair was 69 years later[90].

13484+2337 (LDS 4410) Fragile system with a white dwarf (EGGR 438); CAMELOT image in Fig. 1, right panel.

18180+3846 (GIC 151) Pair with appreciable orbital variation.

19072+2053 (LDS 1017) …

19147+1918 (LDS 1020) Pair of two supposed M3.5V stars (PMSU) with a faint background visual companion at $\rho \sim 3.5$ arcsec to the primary (2MASS J19143845+1919123). The primary is 1.5 mag brighter in $J$ than the secondary and has *Hipparcos* parallactic distance. Oddly, Shkolnik *et al.*[91] classified the primary as a single M4.5V. The scenario that best matches the observables is that the primary is actually an M1V (more massive than listed in Table IV). Low-resolution spectroscopy is needed to confirm this hypothesis.

19169+0510 (LDS 6334) Pair at low Galactic latitude made of two well-known stars, V1428 Aql and vB 10, that have received together almost 600 citations. WDS tabulates only three epochs from 1942 to 1999, and the last one (2MASS) is wrong. We performed a simple astrometric follow-up with public and IAC80 data covering numerous epochs over 70 years and found no orbital variation (Cortés-Contreras et al., in prep.).

19464+3201 (KAM 3) Pair with period shorter than one millenium and appreciable orbital variation. The position angle of the two first astrometric measurements in 1935.77 and 1936.71 by van de Kamp[92] had uncertainties larger than 2 deg.

19510+1025 (J 124) Pair with a very bright F8V-spectral-type primary of Bayer designation (ο Aql). "AC" in WDS; all extra WDS pair candidates are visual, unbound pairs. All astrometric measurements of the pair[93–95] have been affected by the brightness of the primary.

19539+4425 (GIC 159 + MCY 3) Triple system at less than 5 pc with the shortest reduced period in our sample and appreciable orbital variation (primary is a close double).

19566+5910 (GIC 161) …

20446+0854 (LDS 1046) Pair of supposed M1.5V and M3.5V stars (PMSU) with spectro-photometric distance only and with astrometric follow-up. The primary is brighter by 0.47 mag in $J$ than the secondary. We have assumed that the secondary is in turn an equal-brightness, close binary, which makes the system a triple. High-resolution imaging and/or spectroscopy are needed to confirm this hypothesis.

20555-1400 (LDS 6418) …

20568-0449 (LDS 6420) Pair with a white dwarf (EGGR 202).

21011+3315 (LDS 1049) Pair with spectro-photometric distance only.

21148+3803 (AGC 13 AB,AF + JOD 20) Bright quadruple system of Bayer designation (τ Cyg) made of two close binary systems of 0.2–0.6 arcsec separated by 1.5 arcmin.

21161+2951 (LDS 1053) Pair of two supposed M3.5V stars (PMSU) with spectro-photometric distance only. The primary is brighter by 0.85 mag in $J$ than the secondary. We have assumed an m4.5V spectral type for the secondary (in italics in Tables IV and V). Low-resolution spectroscopy is needed to confirm this hypothesis.

21440+1705 (LDS 6358) Fragile pair of supposed M4.0V and M4.5V stars (PMSU). The primary is brighter by 0.77 mag in $J$ than the secondary and has parallactic distance. The scenario that best matches the observables is that the secondary is actually an M5.0V or even M5.5V (less massive than listed in Table IV). Low-resolution spectroscopy is needed to confirm this hypothesis.

22058+6539 (NI 44) Poorly-known pair with astrometric follow-up, spectro-photometric distance only, period shorter than one millenium, and no spectral characterisation of secondary.

22173-0847 (LDS 782 + BEU 22) Triple system (secondary is a close double). The angular separation and position angle of the first astrometric measurement of the wide 'pair', by Luyten[96] in 1920 ($\rho$ = 7 arcsec, $\theta$ = 225 deg), had large uncertainties.

23294+4128 (GIC 93) Triple system with astrometric follow-up and discordant catalogued proper motions. Secondary is a close double with no WDS designation yet.

23573-1259 (LDS 830) Triple system with spectro-photometric distance only. Secondary is a spectroscopic binary of unknown parameters[60]. From Table V and simple spectral-type–absolute-magnitude relations, the secondary cannot be an equal brightness binary.